\documentclass[aps,prl,twocolumn,superscriptaddress,raggedbottom,showpacs,floatfix]{revtex4}
\usepackage{amsmath,amsfonts,amssymb,amsthm}
\usepackage{bm}
\usepackage{color}
\usepackage{graphicx,latexsym}
\usepackage{epstopdf}
\usepackage{comment}

\begin{document} 
\renewcommand{\vec}{\mathbf}
\renewcommand{\Re}{\mathop{\mathrm{Re}}\nolimits}
\renewcommand{\Im}{\mathop{\mathrm{Im}}\nolimits}

\title{Anomalous Coulomb Drag in Electron-Hole Bilayers due to the Formation of Excitons}
\author{Dmitry K. Efimkin}
\affiliation{Joint Quantum Institute and Condensed Matter Theory Center,  University of Maryland, College Park, Maryland 20742-4111, USA}
\author{Victor Galitski}
\affiliation{Joint Quantum Institute and Condensed Matter Theory Center, University of Maryland, College Park, Maryland 20742-4111, USA}
\affiliation{School of Physics, Monash University, Melbourne, Victoria 3800, Australia}

\begin{abstract}
Several recent experiments have reported an anomalous temperature dependence of the Coulomb drag effect in electron-hole bilayers. Motivated by these puzzling data, we study theoretically a low-density electron-hole bilayer, where electrons and holes avoid quantum degeneracy by forming excitons. We describe the ionization-recombination crossover between the electron-hole plasma and exciton gas and calculate both the intralayer and drag resistivity as a function of temperature. The latter exhibits a minimum followed by a sharp upturn at low temperatures in a  qualitative agreement with the experimental observations [see, e.g., J.~A.~Seamons et al., Phys. Rev. Lett. {\bf 102}, 026804
(2009)]. Importantly, the drag resistivity in the proposed scenario is found to be rather insensitive to a mismatch in electron and hole concentrations in sharp contrast to the scenario of electron-hole Cooper pairing. 
\end{abstract}
\maketitle
Coulomb drag effect is a sensitive probe of interactions and collective phases in bilayer systems (see Ref.~[\onlinecite{Rojo,Levchenko}] for a review). In its usual setup, an electric current in the first layer, $I_\mathrm{drive}$, drags charge carriers in the other one. If the second layer is closed, the drag force is compensated by the Coulomb force induced by a voltage drop, $V_\mathrm{drag}$, and the drag resistivity of the bilayer $\rho_\mathrm{D}=V_\mathrm{drag}/I_\mathrm{drive}$ is measured. If the bilayer involves two weakly-coupled Fermi liquids, the temperature dependence of the drag resistivity at low temperatures is quadratic $\rho_\mathrm{D}\sim T^2$, which is well established both theoretically~\cite{JauhoSmith,ZhenMacDonald,KamenevOreg} and experimentally~\cite{DragExp1,DragExp2}. Any deviations from that Fermi-liquid behavior can signal the appearance of collective phases or correlations in the bilayer system.

A number of recent experiments~\cite{ExcitonExp1,ExcitonExp2, ExcitonExp3, ExcitonExp4} on the electron-hole $\hbox{GaAs}/\hbox{GaAlAs}$ bilayers  have observed an anomalous temperature dependence of drag resistivity at at the intermediate doping $n_{e(h)}\approx 5\; 10^{10} \;\mathrm{cm}^{-2}$.  The $T$-dependence of $\rho_\mathrm{D}$ was shown to achieve a minimum, followed by a growth and saturation at lower temperatures, which were rather insensitive to the concentrations mismatch (see also related experiments for other realizations of electron-hole bilayers~\cite{GrapheneExp, QHFExp1,QHFExp2}). This behavior cannot be attributed to interlayer exchange and correlation effects~\cite{HwangDasSarma,BadalyanKimVignaleSenatore,HwangDasSarmaBraudeStern}, that are relevant in that regime, and does not appear for electron-electron and hole-hole bilayers for similar parameters. Therefore, there is strong evidence for an excitonic origin of the effect, but its detailed understanding  is still lacking. 

There were a number of theoretical attempts to explain the experiments based on the Bardeen-Cooper-Schrieffer (BCS) model of  electron-hole Cooper pairing \cite{LozovikYudson1, LozovikYudson2, Shevchenko, MacDonaldRezayi}, which is valid in the high doping regime and can be the origin of the dipolar superfluidity \cite{LozovikYudson1,JoglekarBalatsky, JoglekarBalatskyLilly}. The mean-field theory predicts a jump of drag resistivity at the pairing temperature to a value comparable to a single layer resistivity~\cite{VignaleMacDonald}. The jump can be smoothed by  pairing fluctuations~\cite{EfimkinJosephson,RistVarlamovMacDonaldFazioPolini}, which are a precursor to the paired state, and both Aslamazov-Larkin~\cite{EfimkinDrag} and Maki-Thomson~\cite{Hu,MinkStoofDuinePoliniVignalePRL,MinkStoofDuinePoliniVignalePRB} contributions are important here. However, Cooper pairing and the fluctuations are very sensitive to the mismatch \cite{EfimkinLOFF, Seradjeh} in contrast to experimental observations.   

\begin{figure}[t]
\label{Fig1}
\vspace{-2 pt}
\includegraphics[width=7.7 cm]{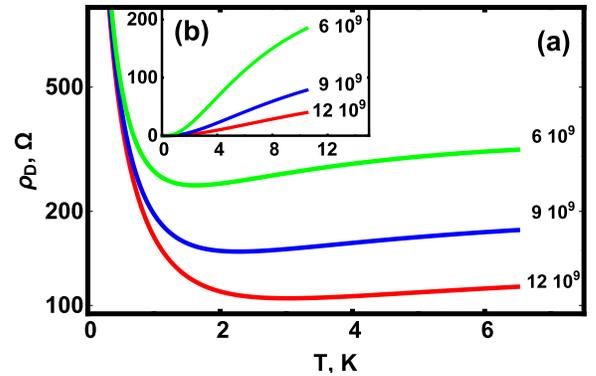}
\vspace{-2 pt}
\caption{(Color online) Shown is the dependence of the drag resistivity $\rho_\mathrm{D}$  on temperature for matched concentrations of electrons and holes with excitons [Fig.~1(a)] and without excitons [inset~1(b)]. The curves correspond to different density per layer $n_\mathrm{e(h)}^0$ denoted by their values in $\hbox{cm}^{-2}$, and axes of the inset coincide with ones of the main plot. The drag resistivity $\rho_\mathrm{D}$ achieves a minimum at $T_\mathrm{D}$ within ionization-recombination crossover between the high-temperature  regime, $T\gg E_\mathrm{exc}$, where the drag is dominated by Coulomb interactions in the  electron-hole plasma, and the low-temperature regime, $T\approx E_\mathrm{exc}$, where the drag is dominated by excitons.}
\end{figure}

Here we present an alternative theoretical scenario for the effect involving the formation of excitons, which are a bound state of spatially separated electrons and holes, with a small binding energy, $E_\mathrm{exc}$. For $T\gg E_\mathrm{exc}$, excitons ionize to form a classical electron-hole plasma and the drag effect is dominated by the Coulomb interactions. At low temperatures, the appearance of excitons strongly enhances the drag and single-layer resistivities, leading to the upturn in the former. The anomalous behavior is robust against the mismatch in the concentration of electrons and holes: while the  magnitude of the upturn is affected by it, the temperature $T_\mathrm{D}$, where the drag resistivity reaches minimum is insensitive to the mismatch. Proposed scenario is valid and self-consistent at low doping, and the calculated excitonic upturn is considerable larger than the observed one. Nevertheless, our results are in a qualitative agreement with the existing experiments.  The main conclusion of our work is that the picture of exciton formation is more relevant to the intermediate doping regime in experiments, than the scenario of electron-hole Cooper pairing.

\emph{Model and the excitonic crossover.} The system of spatially separated electrons and holes, which can bind to form excitons, can be described by the following Hamiltonian
\begin{equation}
\label{Hamiltonian}
\begin{split}
\hat{H}=\sum_{\vec{p}S} \epsilon_{exc}(\vec{p}) b_{\vec{p}S}^+b_{\vec{p}S} + \sum_{\vec{p}s} \epsilon_{\alpha}(\vec{p}) a_{\alpha\vec{p}s}^+a_{\alpha\vec{p}s}+ \\ +
\frac{1}{2} \sum_{\begin{smallmatrix} \vec{p} \vec{p}^\prime\vec{q}\\ s s^\prime \alpha \alpha^\prime \end{smallmatrix}} V_{\alpha\alpha^\prime}(\vec{q}) a^+_{\alpha,\vec{p}+\vec{q},s} a^+_{\alpha^\prime,\vec{p}^\prime-\vec{q},s^\prime} a_{\alpha^\prime\vec{p}^\prime s^\prime} a_{\alpha\vec{p}s}. 
\end{split} 
\end{equation}
Here $a_{\alpha\vec{p}s}$ and $b_{\vec{p}s}$ are annihilation operators for electrons ($\alpha=\mathrm{e}=1$), holes ($\alpha=\mathrm{h}=-1$) and excitons with momentum $\vec{p}$ and internal degeneracy spin index $s=(|\!\downarrow\rangle,|\!\uparrow\rangle)$ and $S=(|\!\downarrow\downarrow\rangle,|\!\uparrow \downarrow\rangle), |\!\downarrow \uparrow \rangle, |\!\uparrow \uparrow \rangle)$. Their dispersions are $\epsilon_{\alpha}(\vec{p})=p^2/2m_\alpha$ and $\epsilon_{\mathrm{exc}}(\vec{p})=p^2/2m_\mathrm{exc}-E_\mathrm{exc}$ with $m_\mathrm{exc}=m_\mathrm{e}+m_\mathrm{h}$ and $E_\mathrm{exc}$ being the exciton mass and its binding energy;  $V_{\alpha\alpha}(\vec{q})=2\pi e^2/\epsilon q$ and $V_{\alpha\bar{\alpha}}(\vec{q})=-2\pi e^2 e^{-q d}/\epsilon q$ are bare intralayer and interlayer Coulomb interactions with interlayer spacing $d$ and bare dielectric permittivity $\epsilon$. We do not specify the interaction with disorder explicitly, but assume relaxation times $\tau_\alpha$ and $\tau_\mathrm{exc}$ to be momentum independent, which implies the short-range disorder to be the dominant scattering mechanism.   

For all numerical calculations we use the set of parameters related to the $\hbox{GaAs}/\hbox{GaAlAs}$ bilayer in experiments~\cite{ExcitonExp1}: $m_\mathrm{e}\approx 0.067 m_0$, $m_\mathrm{h}\approx 0.4 m_0$, $d\approx 30\,\hbox{nm}$, $\epsilon=12.4$ with $m_0$ the bare electronic mass. The relaxation times $\tau_\alpha$ are parametrized by mobilities $M_\mathrm{e}\approx 2\; 10^6 \;\hbox{cm}^2/\hbox{V} \hbox{s}$, and $M_\mathrm{h}\approx 3\; 10^5 \;\hbox{sm}^2/\hbox{V} \hbox{s}$. The excitonic relaxation time, $\tau_\mathrm{exc}=m_* \tau_\mathrm{e} \tau_\mathrm{h}/(\tau_\mathrm{e} m_\mathrm{e}+\tau_\mathrm{h} m_\mathrm{h})$, where the reduced mass is $m_*=m_\mathrm{e}m_\mathrm{h}/(m_\mathrm{e}+m_\mathrm{h})$, corresponds to the mobility $M_\mathrm{exc}\approx 3.4\; 10^4 \;\hbox{cm}^2/\hbox{V} \hbox{s}$. Nevertheless, excitons being nonlocal objects are more sensitive to interlayer tunneling and other factors, so their mobility can be considerably reduced and here we use $M_\mathrm{exc}\approx 10^4 \;\hbox{cm}^2/\hbox{V} \hbox{s}$.
The effective Bohr radius, $a_\mathrm{B}= \hbar^2 \epsilon/e^2 m_* \approx 11.8\; \hbox{nm}$, and Rydberg energy, $E_\mathrm{B}=m_* e^4/2 \hbar^2 \epsilon^2 =55.4\; \hbox{K}$, give the spatial and energy scales. The exciton energy, $E_\mathrm{exc}$, can be considerably smaller than $E_\mathrm{B}$ at $d\gtrsim a_\mathrm{B}$ and is sensitive to screening, so here we use $E_\mathrm{exc}\approx 0.5\; \hbox{K}$, corresponding to the exciton size $a_\mathrm{exc}\approx 110 \; \mathrm{nm}$, as an independent parameter. The model is self-consistent if excitons weakly overlap, which corresponds to the doping $n_\mathrm{e(h)}\lesssim 10^{10}\; \hbox{cm}^{-2}.$  

\begin{figure}
\label{Fig2}
\begin{center}
\includegraphics[width=8.4 cm]{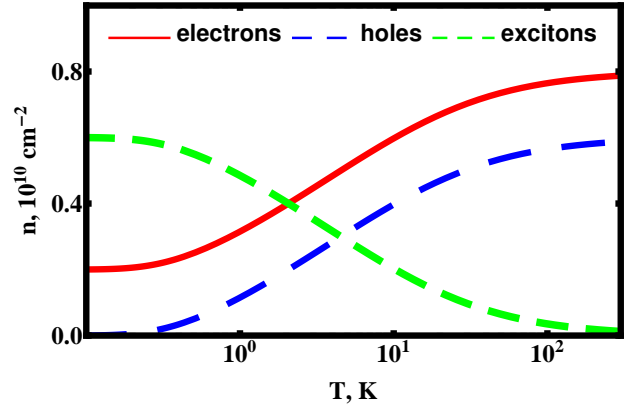}
\vspace{-2 pt}
\caption{(Color online) The temperature dependence of concentrations of electrons $n_\mathrm{e}$, holes $n_\mathrm{h}$ and excitons $n_\mathrm{exc}$, which are given by the Eqs.~(\ref{Concentrations}), for fixed total concentrations per layer $n_\mathrm{e}^0=8\;10^9 \hbox{cm}^{-2}$, $n_\mathrm{h}^0=6\;10^9 \hbox{cm}^{-2}$. The dependencies for other values of $n_{\mathrm{e}(h)}^0$ are qualitatively similar. At high temperatures $T \gg E_\mathrm{exc}$ there is a long excitonic tail $n_\mathrm{exc}\approx n_\mathrm{e}^0 n_\mathrm{h}^0/n_* \sim T^{-1}$. The concentration of excitons at zero tempearture is equal to that of the minority species (holes in this case) in the limit of large temperatures.}
\vspace{-2 pt}
\end{center}
\end{figure}
  
The ground state of the model is believed to be the exciton condensate that forms at the temperature $T_\mathrm{Q}\lesssim E_\mathrm{exc}$ and can coexist with the degenerate gas of electrons or holes in the presence of their concentration mismatch.  However, below we focus on the ionization-recombination crossover regime $T \gtrsim E_\mathrm{exc}$, where the distributions of electrons, holes and excitons are non-degenerate. To calculate their concentrations we recall that in experiments the total concentrations of charged particles per layer $n_\mathrm{\alpha}^0$ are controlled independently by electrical doping, so $n_\mathrm{exc} + n_\mathrm{\alpha}=n_\mathrm{\alpha}^0$. Here $n_\alpha$ and $n_\mathrm{exc}$ are concentrations of quasiparticles. Reintroducing the grand canonical Hamiltonian, 
$\hat{H}_\Omega=\hat{H}-\sum_\alpha \mu_\alpha (\hat{n}_\mathrm{exc} + \hat{n}_\alpha )$, with chemical potentials $\mu_\alpha$ as Lagrange multipliers, we get the chemical potential of excitons as $\mu_\mathrm{exc}=\mu_\mathrm{e}+\mu_\mathrm{h}$. The equation for concentrations can be simplified to
$n_\mathrm{e} n_\mathrm{h}/n_* + n_\mathrm{\alpha}= n_\mathrm{\alpha}^0$, where the concentration $n_*= m^* T\exp[-E_\mathrm{exc}/T]/(2\pi \hbar^2)$. The temperature dependencies of fermionic and excitonic concentrations are given by 
\begin{equation}
\begin{split}
\label{Concentrations}
n_\alpha=\frac{1}{2}\left[\delta n_\alpha^0-n_*+\sqrt{(\delta n_\alpha^0)^2+n_*^2 +2 n_* n_\mathrm{T}^0}\right]; \\
n_\mathrm{exc}=\frac{1}{2} \left[  n_\mathrm{T}^0+n_*-\sqrt{(\delta n_\alpha^0)^2+n_*^2 +2 n_* n_\mathrm{T}^0}\right],
\end{split}
\end{equation}
where $\delta n_\alpha^0=n_\alpha^0-n_{\bar{\alpha}}^0$ and $n_\mathrm{T}^0=n_\mathrm{e}^0+n_\mathrm{h}^0$ are the concentration mismatch and the total concentration. 

The temperature dependence of the concentrations is depicted in Fig.~2. At low temperatures $T\ll E_\mathrm{exc}$ the fraction of unbound electrons and holes is exponentially small, while within the crossover $T\gtrsim E_\mathrm{exc}$ there is a long non-degenerate tail of excitons decreasing as $T^{-1}$ according to $n_\mathrm{exc}\approx n_\mathrm{e}^0 n_\mathrm{h}^0/n_*$. The exciton gas can be considered  non-degenerate until $T_\mathrm{Q}\approx 0.3\, \hbox{K}$~\cite{Comment1}. 

\emph{Phenomenology of the drag effect.} In the presence of electrons, holes and excitons the conductivity tensor of the bilayer system is given by 
\begin{equation}
\label{ConductivityTensor}
\left(
\begin{array}{cc}
J_{\mathrm{e}}\\
J_{\mathrm{h}}\\
\end{array}\right)=\left(
\begin{array}{cc}
\sigma_{\mathrm{exc}} + \sigma_{\mathrm{e}} & - \sigma_{\mathrm{exc}}- \sigma_{\mathrm{D}}\\ -\sigma_{\mathrm{exc}} - \sigma_{\mathrm{D}}& \sigma_{\mathrm{exc}} + \sigma_{\mathrm{h}}\\  \end{array}\right) \left(\begin{array}{cc}
E_{\mathrm{e}}\\
E_{\mathrm{h}}\\
\end{array}\right),
\end{equation}
where $\sigma_\alpha=n_\alpha e^2 \tau_\alpha/m_\alpha$ and  $\sigma_\mathrm{exc}=n_\mathrm{exc} e^2 \tau_\mathrm{exc}/m_\mathrm{exc}$ are their Drude conductivities. Excitons, being composed of electrons and holes from different layers, contribute to both diagonal and off-diagonal components of the conductivity tensor with opposite signs. The transconductivity $\sigma_\mathrm{D}$ originates from the Coulomb interaction between electrons and holes and is calculated microscopically below. The drag resistivity $\rho_\mathrm{D}$ and single layer resistivities $\rho_\alpha$, being the components of the inverted conductivity matrix (\ref{ConductivityTensor}), can be written in a compact way 
\begin{equation}
\label{Transresistivity1}
\rho_{\mathrm{D}(\alpha)}=\frac{\sigma_{\mathrm{D}(\bar{\alpha})}+ \sigma_\mathrm{exc}}{\sigma_\mathrm{e}\sigma_\mathrm{h} + (\sigma_\mathrm{e}+\sigma_\mathrm{h})\sigma_\mathrm{exc}}.
\end{equation}
At zero temperature the excitonic contribution dominates and they become 
\begin{equation}
\label{Transresistivity2}
\begin{split}
\rho_\mathrm{D}=\sum_{\alpha} \frac{\Theta_{\alpha \bar{\alpha}} m_\alpha}{(n_\alpha^0-n_{\bar{\alpha}}^0) e^2 \tau_\alpha}; \quad
\rho_\alpha=\rho_\mathrm{D}+ \frac{\Theta_{\bar{\alpha}\alpha } m_\mathrm{exc}}{n_\alpha^0 e^2 \tau_\mathrm{exc}}.
\end{split}
\end{equation}
Here $\Theta_{\alpha\bar{\alpha}} =\Theta(n_\alpha-n_{\bar{\alpha}})$ is the Heaviside function. If densities of electrons and holes are perfectly matched, both single layer resistivities $\rho_\alpha$ and  $\rho_\mathrm{D}$ diverge at $T=0$. This corresponds to an insulating excitonic ground state with the perfect drag effect:  the relation between the electric current in a layer, induced by a current in the other layer, is $I_\mathrm{drag}=-I_\mathrm{drive}$.  Our considerations assume $T\gg T_\mathrm{Q}$, where there is a competition between $\sigma_\mathrm{exc}$ and $\sigma_\mathrm{D}$, but the zero-temperature values (\ref{Transresistivity2})  reflect the strength of the low-temperature upturn.
 
\emph{Electron-hole transconductivity.} The transconductivity $\sigma_\mathrm{D}$ can be calculated in the second order of perturbation theory in the interlayer Coulomb interaction \cite{KamenevOreg} as follows
\begin{equation}
\begin{split}
\label{SigmaDrag}
\sigma _\mathrm{D}=-\frac{1}{16 \pi T} \sum_{\vec{q}}\int_{-\infty}^\infty  \frac{d \omega}{\sinh^2(\frac{\omega}{2T})}\Gamma_{x \mathrm{e}}^\mathrm{RA}(\vec{q},\omega,\omega) \times \\ \Gamma_{x\mathrm{h}}^\mathrm{AR} (\vec{q},\omega,\omega)|U_\mathrm{eh}(\vec{q},\omega)|^2,
\end{split}
\end{equation}
where $U(q,\omega)$ is the screened interlayer interaction and $\Gamma_{x \alpha}^\mathrm{RA}(\vec{q},\omega,\omega)$ is the current-charge-charge nonlinear susceptibility. If the relaxation times $\tau_\alpha$ are momentum independent, as we assume here, it is given by \cite{FlensbergHu2}
\begin{equation} 
\label{GammaVertex}
\Gamma_{x \alpha}^\mathrm{RA}(\vec{q},\omega,\omega)=\alpha q_x \frac{e \tau_\alpha}{m_\alpha} \Pi_{\alpha 2}^\mathrm{R}(\vec{q}, \omega),
\end{equation} 	
where $\Pi_{\alpha\mathrm{2}}^\mathrm{R}(\vec{q}, \omega)$ is the imaginary part of the polarization operator, which for a non-degenerate gas is given by~\cite{SM}
\begin{equation}
\label{POIm}
\Pi_{\alpha 2}^\mathrm{R}=-\frac{ \sqrt{\pi}n \tilde{q}_\alpha}{T q} \sinh\left[\frac{\omega}{2 T}\right] \exp\left[-\frac{\tilde{q}_\alpha^2 \omega^{2}}{4 T^2 q^{2}} -\frac{q^2}{4 \tilde{q}_\alpha^{2}}\right].
\end{equation}  
Here the $\tilde{q}_\alpha =\sqrt{2 m_\alpha T}$ is the characteristic thermal momentum scale. For the interaction $U(\vec{q},\omega)$, the static Debye-H$\ddot{\hbox{u}}$ckel approximation, that ignores the presence of neutral excitons, yields
\begin{equation}
\label{InterlayerInteraction}
U_\mathrm{eh}(\vec{q})=\frac{2 \pi e^2}{\epsilon} \frac{q e^{-q d}}{(q+\kappa_\mathrm{\mathrm{e}}) (q+\kappa_\mathrm{\mathrm{h}})- \kappa_\mathrm{\mathrm{e}} \kappa_\mathrm{\mathrm{h}} e^{-2 qd}}.
\end{equation}
Here $\kappa_\alpha(\vec{q})=\kappa_\alpha^0 f_\kappa (q/\tilde{q}_\alpha)$, $\kappa_\alpha^0=2\pi e^2 n_\alpha/\epsilon T$ is the Debye-H$\ddot{\mathrm{u}}$ckel screening momentum and $f(x)$ is the dimensionless function $f(2 x)=\sqrt{\pi}\exp[-x^2] \mathrm{Erfi}(x)/2x$ with $\mathrm{Erfi}(x)$ to be the imaginary error function. The static screening approximation does not take into account possible plasmon contribution~\cite{FlensbergHu1, FlensbergHu2}, which considerably enhances the drag effect for $0.4\lesssim T/\mu \lesssim 1$. However, in the non-degenerate regime, the plasmons 	become strongly damped, and can be ignored. The integral over frequencies in Eq.~(\ref{SigmaDrag}) can be calculated explicitly and we get
\begin{equation}
\label{SigmaDrag2}
\sigma_\mathrm{D}=\frac{\sqrt{\pi}}{32} \frac{e^2}{h} \frac{\tau_\mathrm{e} \tau_\mathrm{h}}{\hbar^2} \frac{q_\mathrm{d}^4}{m_\mathrm{e} m_\mathrm{h}} I_q 
\end{equation}
with momenta $q_\mathrm{d}=\hbar d^{-1}$, $\tilde{q}_*=\sqrt{2 m^* T}$  and a dimensionless integral $I_q$ over rescaled momentum $\vec{q}$ given by
\begin{equation}
\label{Iq}
I_q=\int_0^\infty dx \frac{ \tilde{q}_* \kappa_\mathrm{\mathrm{e}}^0 \kappa_\mathrm{\mathrm{h}}^0 d^3 \; \; x^4 e^{-2x} e^{-\frac{x^2}{4 \tilde{q}_*^2 d^2}} }{[(x+\kappa_\mathrm{\mathrm{e}}d) (x+\kappa_\mathrm{\mathrm{h}}d)- \kappa_\mathrm{\mathrm{e}} \,\kappa_\mathrm{\mathrm{h}}d^2 e^{-2 x}]^2}.
\end{equation}
There are three different momenta $q_\mathrm{d}$, $\tilde{q}_*$, $\kappa_\alpha^0$ (for calculations of asymptotes we assume that $\kappa_\mathrm{e}^0$ and $\kappa_\mathrm{h}^0$ have the same order of magnitude) in the integral $I_q$, and the characteristic momentum, transfered between electron and hole layers, is the smallest of them. If these momenta are well separated, the asymptotic behavior of the integral $I_q$ can be evaluated analytically. There are four different regimes: $I:\tilde{q}_*\ll q_\mathrm{d}, \kappa_\alpha^0$; $II_+: q_\mathrm{d} \ll \tilde{q}_*, \kappa_\alpha^0$; $II_-: \kappa_\alpha^0 \ll \tilde{q}_* \ll q_\mathrm{d}$ and $III: \kappa_\alpha^0\ll q_\mathrm{d}\ll \tilde{q}_*$ with
\begin{equation}
\label{Assymptotics}
\begin{split}
I: \; I_q= \frac{\sqrt{\pi}}{2} \frac{\tilde{q}_*^4 d^2}{\kappa_\mathrm{e}^0 \kappa_\mathrm{h}^0}; \quad 
II_+:\; I_q= \frac{\pi^4}{120} \frac{\tilde{q}_* d^{-1}}{\kappa_\mathrm{e}^0 \kappa_\mathrm{h}^0};  \\ 
II_-: \;I_q = \sqrt{\pi} \tilde{q}_*^2 \kappa_\mathrm{e}^0 \kappa_\mathrm{h}^0 d^4; \quad
III:\; I_q=\frac{\tilde{q}_* \kappa_\mathrm{e}^0 \kappa_\mathrm{h}^0 d^3}{2}.  
 \end{split} 
\end{equation}
Regimes $I$ ($T<T_1^\pm$) and $III$ ($T_2^\pm<T$) appear at small and large temperatures. Depending on the concentration $n_\mathrm{e(h)}$ one of $II_-$ and $II_+$ is between them ($T_1^\pm<T<T_2^\pm$).   
The corresponding boundaries are given by $T_2^+=4\pi E_\mathrm{B} n_\alpha a_\mathrm{B} d$, $T_2^-=T_1^+= E_\mathrm{B} (a_\mathrm{B}/d)^2$ and $T_1^-=E_\mathrm{B}(4\pi n_\alpha a_\mathrm{B}^2)^{2/3}$. The point at which $T_1^\pm=T_2^\pm$ and the regimes $II_\pm$ merge and disappear corresponds to $n_{12}=a_\mathrm{B}/4\pi d^3=3.1 \; 10^9\,\;\hbox{cm}^{-2}$ and $T_{12}=E_\mathrm{B} (a_\mathrm{B}/d)^2\approx 7.1\;\hbox{K}$. For the densities of interest, the momentum scales are not well separated, the range of the applicability of the asymptotes (\ref{Assymptotics}) is reduced to $T\ll T_1^\pm$ and $T\gg T_2^\pm$. Below, we calculate $I_q$ numerically.  
\begin{figure}
\label{Fig3}
\begin{center}
\includegraphics[width=8.4 cm]{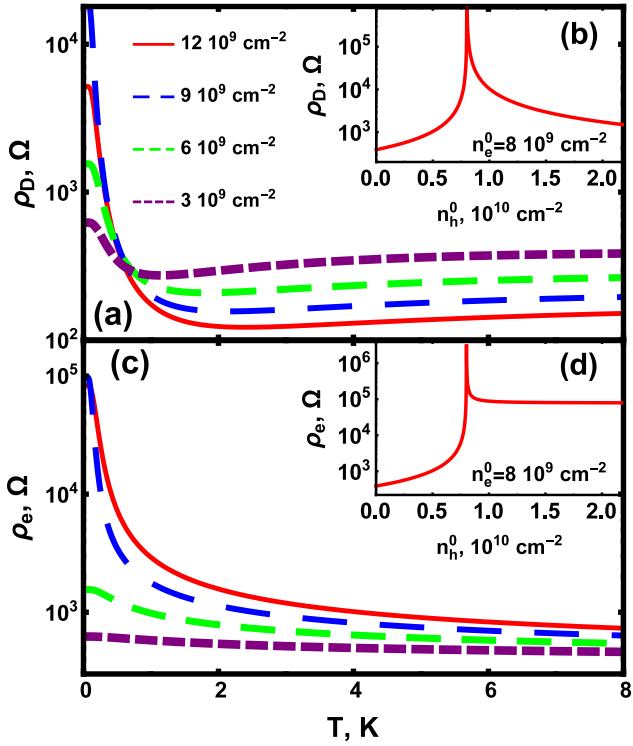}
\vspace{-2 pt}
\caption{(Color online) (a) and (c) The temperature dependencies of the drag resistivity $\rho_\mathrm{D}$ and the resistivity of electrons $\rho_\mathrm{e}$ in the presence of the mismatch in electron and hole concentrations. (b) and (d) The corresponding values at zero temperature, which are given by Eqs.~(\ref{Transresistivity2}). The strength of the excitonic enhancement of both $\rho_\mathrm{D}$ and $\rho_\alpha$ is defined by the mismatch, while the temperature dependencies are quite insensitive to it.}
\vspace{-2 pt}
\end{center}
\end{figure}
\emph{Drag resistivity of the bilayer.} First, it is instructive to analyze the dependence of  drag resistivity $\rho_\mathrm{D}$ on the temperature $T$ ignoring the presence of excitons. It is shown in Fig.~1-b (for matched concentrations of electrons and holes). At high temperatures $T_2^+ \lesssim T$, the screening disappears, making $\kappa_\alpha^0$ the smallest momentum scale, and the drag resistivity decreases as $\rho_\mathrm{D}\sim T^{-3/2}/d$. In the intermediate regime $T_{1}^+ \lesssim T \lesssim T_{2}^+$, the scattering momentum is $q_\mathrm{d}$ and the asymptotic form is $\rho_\mathrm{D}\sim T^{5/2}/n_\mathrm{e}^2 n_\mathrm{h}^2 d^5$. These two scattering regimes are usual for bilayer fermion systems (along with regimes where plasmons \cite{FlensbergHu1, FlensbergHu2} and phonons \cite{BosnagerFlensbergHuMacDonald} dominate and the hydrodynamic one~\cite{ApostolovLevchenkoAndreev, ChenAndreevLevchenko}), but the latter corresponds to $\rho_\mathrm{D}\sim T^2$ due to degeneracy of fermions. For the considered system, at low temperatures $T\lesssim T_{1}^+$, the electrons and holes avoid degeneracy by transforming into excitons and their characteristic momentum scale $\tilde{q}_*$ becomes the scattering one leading to the asymptotic behavior $\rho_\mathrm{D}\sim T^4/n_\mathrm{e}^2 n_\mathrm{h}^2 d^2$. That regime usually does not appear in a fermionic bilayer due to quantum degeneracy of fermions.

The temperature dependence of   $\rho_\mathrm{D}$ in the presence of excitons  are presented in Fig.~1-a (perfectly matched densities) and Fig.~3-a (with a mismatch).  The latter is supplemented by the inset Fig.~3-b in which the dependence of $\rho_\mathrm{D}$ on the mismatch  at zero temperature is depicted. The long excitonic tail, which weakly depends on temperature, considerably enhances the drag resistivity even at high temperatures $T\gg E_\mathrm{exc}$. The dependence has a clear minimum at the temperature $T_\mathrm{D}$, which lies within the crossover $E_\mathrm{exc}\lesssim T_\mathrm{D} \lesssim T_2^\pm$. The strength of the upturn is defined by the mismatch, while the temperature $T_\mathrm{D}(n_\mathrm{e}^0, n_\mathrm{h}^0)$ smoothly increases with both its arguments and does not have any features for the matched case. This makes the minimum in the temperature dependence of $\rho_\mathrm{D}$ shallower with increasing of both concentrations, as seen in the experiment. The excitonic contribution to the drag resistivity $\rho_\mathrm{D}$ can be well-fitted by a combination of functions $T^{-1}$ and $T^{-2}$. The former dominates at high temperatures $T\gg E_\mathrm{exc}$, while the latter plays the major role at $T\sim E_\mathrm{exc}$. At lower temperatures the drag resistivity saturates to a value, which depends on the imbalance of concentrations (see Fig.3-b).   

Resistivity of electrons is presented in Fig.3-c and supplemented by the inset (d), where its dependence on the mismatch at $T=0$ is depicted. Depending on the mismatch, its enhancement vary by an order of magnitude, while the temperature dependence is quite insensitive to it. The dependencies for the resistivity of holes are qualitatively the same.  
 
\emph{Discussion.} The proposed scenario of genuine excitonic drag effect does not assume any phase transition and/or coherence of excitons, which in our model may occur at lower temperatures $T_\mathrm{Q}$ (Localization effects and their interplay with other ground states, not involving exciton condensation, can not be ruled out: e.g., an excitonic Bose glass~\cite{BoseGlass,BoseGlassExciton} or an exotic Bose-metal phase~\cite{BoseMetal}, which was conjectured to exist in models involving dirty composite bosons and gapless fermionic excitations). We argue however that the upturn in $\rho_\mathrm{D}$ is unrelated to the quantum effects including localization, but appears at the temperature $T_\mathrm{D}$ corresponding to the ionization-recombination excitonic crossover $E_\mathrm{exc}\lesssim T_\mathrm{D} \lesssim T_2^\pm$ from a classical electron-hole plasma to a classical exciton gas. The exact value of the $T_\mathrm{D}$ is non-universal and depends on the interlayer distance, quasiparticle mobilities, effective masses, etc.  


For explicit calculations above, we have used a range of electron and hole concentrations, which is about an order of magnitude smaller than the ones in the published experiments to ensure that the assumptions of our model are self-consistent. In the intermediate doping regime realized in experiment so far, the excitons overlap and cannot be considered as two-particle objects anymore. To develop a quantitative many-body theory for the Coulomb drag effect in this intermediate regime is difficult, because of complicated interplay of the Pauli blocking effects, self-consistent screening, and coexistence of excitons with degenerate gas of electrons and holes. The extrapolation of our results to this regime considerably overestimates the strength of the excitonic upturn seen in experiments. Nevertheless, the observed behavior of the drag resistivity on temperature and concentrations is qualitatively captured, and we conclude that the picture of exciton formation is more relevant to the experiments, than the scenario of electron-hole Cooper pairing and pairing fluctuations.

\emph{Acknowledgement.} This work was supported by the DOE-BES (Grant No. DESC0001911) and the Simons Foundation. D.K.E is grateful to Yuri Lozovik for turning his attention to the considered problem.

\bibliographystyle{apsrev}
\bibliography{ExcitonBibliography}

\begin{thebibliography}{42}
\bibitem[{Com()}]{Comment1}
\bibinfo{note}{The measure of degeneracy is the fugacity
  $z_{\alpha(\mathrm{exc})}=\exp[\mu_{\alpha(\mathrm{exc})}/T]$. We have
  explicitly checked that along the crossover $z_\alpha(n_\alpha)\lesssim 1$
  and $z_\mathrm{exc}(n_\mathrm{exc})\lesssim 1$, while $z_\alpha(n_\alpha^0)
  \approx 1$, which means that electrons and holes avoid degeneracy by binding
  into excitons. Excitons become degenerate at $T_\mathrm{Q}\approx 0.3 \;
  \mathrm{K}$.}
\end{thebibliography}

\widetext
\clearpage
\begin{center}
\textbf{\large Supplemental Material: "Anomalous Coulomb Drag in Electron-Hole Bilayers due to the Formation of Excitons" by Dmitry K. Efimkin and Victor Galitski}
\end{center}
\setcounter{equation}{0}
\setcounter{figure}{0}
\setcounter{table}{0}
\setcounter{page}{1}
\makeatletter
\renewcommand{\theequation}{S\arabic{equation}}
\renewcommand{\thefigure}{S\arabic{figure}}
\renewcommand{\bibnumfmt}[1]{[S#1]}
\renewcommand{\citenumfont}[1]{S#1}

The Supplemental Material presents a calculation of the polarization operator $\Pi(\vec{q},\omega)$ of two-dimensional nondegenerate electron gas. The general expression for the polarization operator $\Pi(\vec{q},\omega)$ is given by 
\begin{equation}
\Pi(\vec{q},w)=2 \sum_{\vec{p}}\frac{n_\mathrm{F}(\xi_\vec{p})-n_\mathrm{F}(\xi_{\vec{p}+\vec{q}})}{\xi_\vec{p}+\omega- \xi_{\vec{p}+\vec{q}}+ \bm{i} \delta}.
\end{equation}
Here $\xi_\vec{p}=p^2/2m-\mu$ is the dispersion law of fermions and $n_\mathrm{F}(\xi_\vec{p})\approx \exp[-\beta\xi_\vec{p}]$ is their Maxwell distribution function with chemical $\mu=T \ln (\pi \hbar^2 n/m T)$, where $n$ is the total concentration of fermions including spin degeneracy. It is instructive to calculate the imaginary part of the polarization operator $\Pi_2(\vec{q},\omega)$ at first and it can be presented as follows  
\begin{equation}
\Pi_2(q,w)=-\pi\frac{n}{T} (1-e^{-\beta \omega}) \int_0^\infty \frac{p dp}{m}e^{-\frac{p^2}{2 m T}}\int_0^{2\pi}\frac{d \phi}{2\pi}\delta\left(\omega -\frac{q^2}{2m} - \frac{pq}{m}\cos\left(\phi\right)\right).
\end{equation}
The character energy and momentum of nondegenerate electron gas are $\omega_\mathrm{T}=T$ and $q_\mathrm{T}=\sqrt{2 m T}$, and the polarization operator, which has the dimension of the density of states, has the scale $n/T$. As a result, introducing dimensionless units $p=q_\mathrm{T} \sqrt{x}$, $q=q_\mathrm{T} \bar{q}$, $\omega=\omega_\mathrm{T} \bar{\omega}$, and $\Pi(q,\omega)=-n\bar{\Pi}(\bar{q},\bar{\omega})/T$ we get 
\begin{equation}
\bar{\Pi}_2(\bar{q},\bar{\omega})=\pi (1-e^{-\bar{\omega}})\int_0^\infty dx e^{-x} \int_0^{2\pi} \frac{d \phi}{2 \pi} \delta\left(\bar{\omega}-\bar{q}^2- 2 q \sqrt{x} \cos{\phi} \right).
\end{equation}
The argument of the delta function achieves zero only if $x>x_0$, where $x_0=(\bar{\omega}-\bar{q}^2)^2/4\bar{q}^2$, that leads to 
\begin{equation}
\bar{\Pi}_2(\bar{q},\bar{\omega})= \frac{(1-e^{-\bar{\omega}})}{4\bar{q}} \int_{x_\mathrm{0}}^\infty dx \frac{e^{-x}}{\sqrt{x}} \int_0^{2\pi} d\phi \delta\left(\sqrt{\frac{x_0}{x}}-\cos(\phi)\right)= \frac{(1-e^{-\bar{\omega}})}{2\bar{q}}\int_{x_\mathrm{0}}^\infty \frac{ dx e^{-x}}{\sqrt{x-x_0}}=\sqrt{\pi}\frac{(1-e^{-\bar{\omega}})}{2\bar{q}}e^{-x_0},   
\end{equation}
and finally we get the expression (\ref{POIm}) from the main part of the paper, written in dimensionless units 
\begin{equation}
\bar{\Pi}_2(\bar{q},\bar{\omega})=\frac{\sqrt{\pi}}{q} \sinh\left(\frac{\bar{\omega}}{2}\right)\exp\left[-\frac{\bar{\omega}^2}{4 \bar{q}^2} - \frac{\bar{q}^2}{4}\right].  
\end{equation}
The dependence $\bar{\Pi}_2(\bar{q},\bar{\omega})$ on momentum and frequency is presented in Fig. S1-a. The value $\bar{\Pi}_2(\bar{q},\bar{\omega})$ is nonzero for arbitrary momentum and frequency and has prominent maximum at $\bar{w}\sim 1$ and $\bar{q}\sim 1$, which correspond to the thermal frequency $\omega_\mathrm{T}$ and momenta $q_\mathrm{T}$. The calculated dependence for nondegenerate electron gas drastically differs from one at zero temperature. The value $\bar{\Pi}_2(\bar{q},\bar{\omega})$ at zero temperatures is nonzero only within the continuum of two-particle excitations, which has sharp boundaries. 

In the paper we use the static screening approximation and need only real part of the polarization operator $\bar{\Pi}_1(\bar{q},0)$ in the static limit. It can be calculated from the imaginary part with help of Kramers-Kronig relations as follows 
\begin{equation}
\bar{\Pi}_1(\bar{q},0)= \int_{-\infty}^\infty \frac{d \omega}{\pi} \frac{\bar{\Pi}_2(\bar{q},\bar{\omega})}{\bar{\omega}}= \frac{e^{-\frac{q^2}{4}}}{\sqrt{\pi} q} \int_{-\infty}^\infty \frac{d \bar{\omega}}{\bar{\omega}} \sinh(\frac{\bar{\omega}}{2}) \exp[\frac{\bar{\omega}^2}{4 \bar{q}^2}] = \frac{\sqrt{\pi}}{\bar{q}} \exp\left[-\frac{\bar{q}^2}{4}\right] \mathrm{Erfi}\left[\frac{\bar{q}}{2}\right]\approx\frac{6}{3 \bar{q}^2 -2 q^{\frac{3}{2}}+6}.
\end{equation}
Here $\mathrm{Erfi}(\bar{q}/2)$ is the imaginary error function. The approximate expression interpolates the corresponding asymptotics and well describes the momentum dependence of $\bar{\Pi}_1(\bar{q},0)$ at arbitrary momentum. The momentum dependence of $\bar{\Pi}_1(\bar{q},0)$ is presented in Fig. S1-b. 
\begin{figure}[b]
\label{FigS}
\begin{center}
\includegraphics[width=7.0 cm]{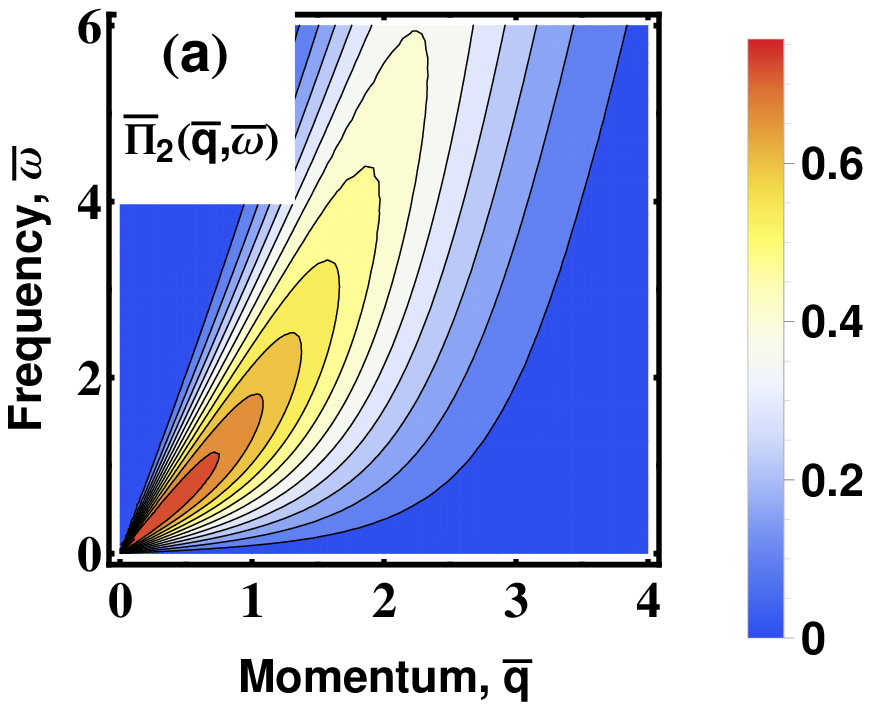}
\includegraphics[width=7.0 cm]{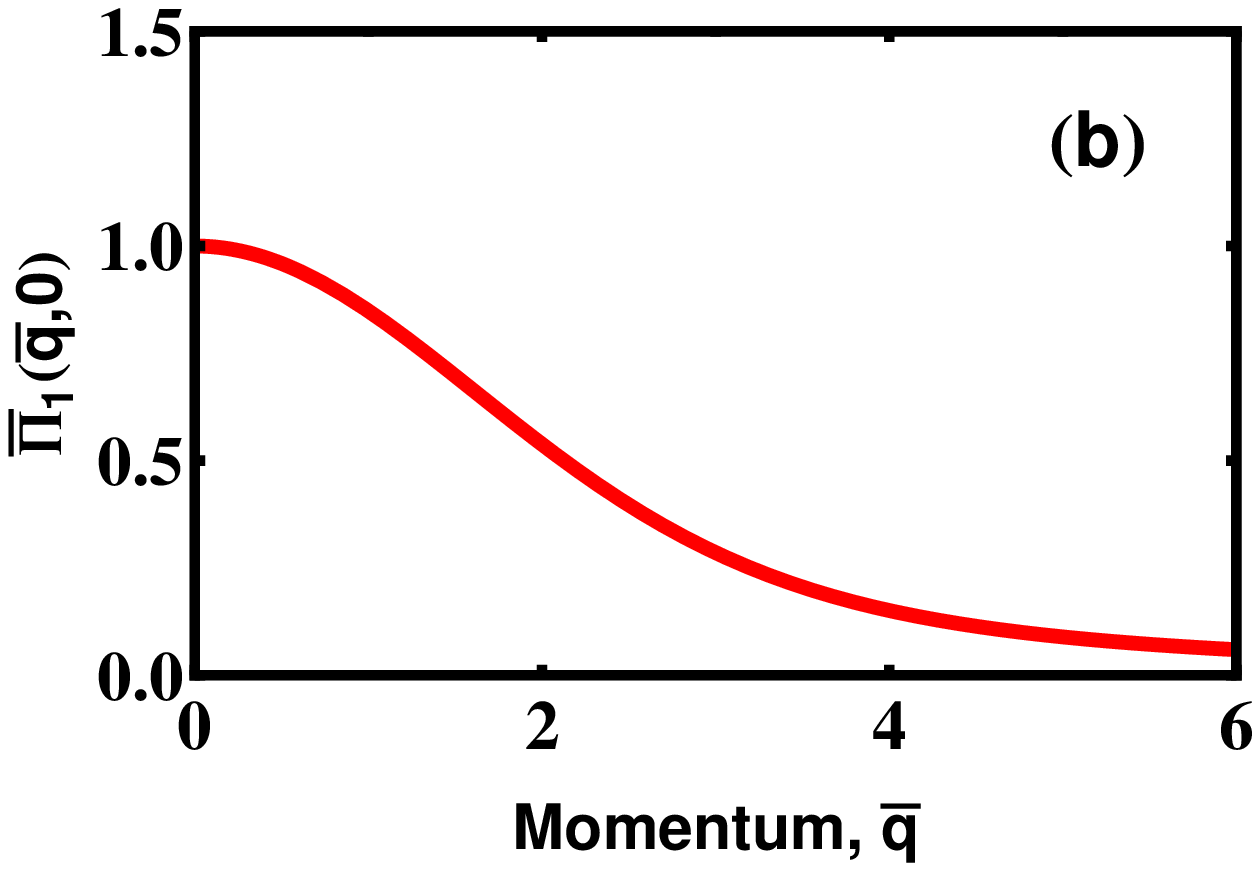}
\vspace{-4 pt}
\caption{a) Frequency and momentum dependence of the imaginary part of polarization operator $\bar{\Pi}_2(\bar{q},\bar{\omega})$; b) Momentum dependence of the real part of the polarization operator $\bar{\Pi}_1(\bar{q},\bar{\omega})$ in the static limit.}
\vspace{-4 pt}
\end{center}
\end{figure}

\end{document}